\documentclass[twocolumn,showpacs,showkeys,reprint,amsmath,amssymb,aps,prl]{revtex4-1}

\usepackage{graphicx}
\usepackage{dcolumn}
\usepackage{bm}
\usepackage{tabularx}
\usepackage{amsmath}

\begin{document}

\title{Ultralow-temperature thermal conductivity of the Kitaev honeycomb magnet $\alpha$-RuCl$_3$ across the field-induced phase transition}

\author{Y. J. Yu,$^{1,\dag}$ Y. Xu,$^{1,\dag}$ K. J. Ran,$^{2,\dag}$ J. M. Ni,$^1$ Y. Y. Huang,$^1$ J. H. Wang,$^2$ J. S. Wen,$^{2,3,\ddag}$ and S. Y. Li$^{1,3,*}$}

\affiliation
{$^1$State Key Laboratory of Surface Physics, Department of Physics, and Laboratory of Advanced Materials, Fudan University, Shanghai 200433, China\\
 $^2$National Laboratory of Solid State Microstructures and Department of Physics, Nanjing University, Nanjing 210093, China\\
 $^3$Collaborative Innovation Center of Advanced Microstructures, Nanjing 210093, China
}

\date{\today}

\begin{abstract}
Recently, there have been increasingly hot debates on whether there exists a quantum spin liquid in the Kitaev honeycomb magnet $\alpha$-RuCl$_3$ in high magnetic field. To investigate this issue, we perform the ultralow-temperature thermal conductivity measurements on the single crystals of $\alpha$-RuCl$_3$ down to 80 mK and up to 9 T. Our experiments clearly show a field-induced phase transition occurring at $H_c$ $\approx$ 7.5 T, above which the zigzag magnetic order is completely suppressed. The minimum of thermal conductivity at 7.5 T is attributed to the strong scattering of phonons by the magnetic fluctuations. Most importantly, above 7.5 T, we do not observe any significant contribution of thermal conductivity from gapless magnetic excitations, which puts a strong constraint on the nature of the high-field phase of $\alpha$-RuCl$_3$.
\end{abstract}

\maketitle

In transition-metal compounds with partially filled $4d$ or $5d$ shells, various novel quantum phases of matter can be brought onto stage, depending on the delicate balance (or the disruption of it) among crystal-field effects, spin-orbit coupling (SOC), and electronic correlation \cite{Witczak}. An incarnation of this competition is the spin-orbit assisted Mott insulator, in which with the help of an enhanced SOC, a Mott gap opening and the formation of $j$$\rm_{eff}$ = 1/2 pseudospins are allowed even for relatively moderate electronic correlation of the $4d$ and $5d$ compounds \cite{SOC assisted1,SOC assisted2}. For such $j$$\rm_{eff}$ = 1/2 Mott insulators on a two-dimensional (2D) honeycomb lattice, unconventional magnetism may be achieved if bond-directional, anisotropic exchange interactions described by a Kitaev model are introduced \cite{Kitaev}. The Kitaev model is one of a few exactly solved 2D models. It supports the long-sought exotic quantum spin liquid (QSL) ground state. Intensively studied hereof are the $5d$ iridates Na$_2$IrO$_3$ and $\alpha$-Li$_2$IrO$_3$ \cite{iridates1,iridates2,iridates3,iridates4}, and more recently the $4d$ analogue $\alpha$-RuCl$_3$ \cite{Kitaev candidate,SOC strength,Nagler,Sears,Cao,Kindo,Baenitz,Park,Raman1,Raman2,neutron Science,Wen,Nagler1,RDJohnson,Lee,Kim,Buchner,Hess,Yu,Buchner1,Klanjsek,Saitoh}.

Despite its much smaller SOC strength ($\lambda \sim 0.1$ eV) compared to iridates, $\alpha$-RuCl$_3$ is a promising candidate for the realization of Kitaev physics and searching for a QSL state therein \cite{Kitaev candidate,SOC strength,Nagler}. $\alpha$-RuCl$_3$ consists of van der Waals coupled layers of edge-sharing RuCl$_6$ octahedra with the central $4d$$^5$ Ru$^{3+}$ ions forming an almost ideal honeycomb lattice \cite{Kitaev candidate}. In iridates, the QSL ground state is preempted by long-range magnetic order, which results from perturbations to the pure Kitaev Hamiltonian due to structural distortions of the IrO$_6$ octahedra \cite{Kitaev candidate}. Although the ground state exhibits a zigzag magnetic order \cite{Nagler,Sears,Cao}, $\alpha$-RuCl$_3$ is free from the structural distortions, so that the residual interactions are smaller, as evidenced by its lower N\'{e}el temperature $T_N$ \cite{Sears,Cao,Kindo,Baenitz,Park}. Indeed, features characteristic of a QSL state have been observed in $\alpha$-RuCl$_3$, such as a broad continuum of magnetic excitations identified in both Raman scattering \cite{Raman1,Raman2} and inelastic neutron scattering measurements \cite{Nagler,neutron Science,Wen,Nagler1}.

To access the potential QSL state in $\alpha$-RuCl$_3$, a reasonable practice would be to suppress the magnetic order by external tuning parameters, such as the magnetic field or pressure. When applying a magnetic field in the honeycomb plane, the zigzag magnetic order is fully suppressed by $H_c$ ($\sim$ 8 T, the value varies slightly in different studies), and the high-field phase after the destruction of the zigzag magnetic order has become the research hotspot recently \cite{Kindo,Baenitz,RDJohnson,Lee,Kim,Buchner,Hess,Yu,Buchner1,Klanjsek}. As for the magnetic excitations in the high-field phase, some studies indicate a gapped scenario \cite{Buchner,Buchner1,Hess,Klanjsek,Kim}, while others suggest a gapless one \cite{Lee,Yu}.

Thermal conductivity measurements have proven to be a powerful technique in probing the elementary excitations in QSL candidates \cite{Matsuda,Matsuda1,Li}. There have already been three works on the thermal conductivity $\kappa$ of $\alpha$-RuCl$_3$ \cite{Lee,Hess,Saitoh}. The high-temperature thermal conductivity exhibits a broad peak around 110 K, which was attributed to itinerant spin excitations due to Kitaev couplings \cite{Saitoh}. Leahy $et$ $al$. reported that the $\kappa$ shows a striking enhancement with linear growth above the critical field $H_c \sim$ 7 T, which was explained as the behavior of proximate Kitaev excitations (PKE) \cite{Lee}. More recently, the in-plane and $c$-axis thermal conductivity are found to show similar behavior in magnetic field, which suggests that the unusual magnetic field dependence is the result of severe scattering of phonons off putative Kitaev-Heisenberg excitations, i.e., of phononic origin \cite{Hess}. The measurements in all three works are conducted at a relatively high temperature range, so that information about the asymptotic behaviors of $\kappa/T$ at $T$ $\rightarrow$ 0, which is crucial for the understanding of the ground state, cannot be obtained.

In this Letter, we report the ultralow-temperature thermal conductivity measurements on a high-quality $\alpha$-RuCl$_3$ single crystal down to 80 mK. A field-induced phase transition is clearly resolved at $H_c$ $\sim$ 7.5 T. Moreover, in the high-field phase of $\alpha$-RuCl$_3$, no significant contribution to $\kappa$ from magnetic excitations is detected. Instead, we find that magnetic excitations only affect the phonon thermal conductivity by scattering. We will discuss the implications of our findings on the high-field phase of $\alpha$-RuCl$_3$.

Single crystals of $\alpha$-RuCl$_3$ were grown by the chemical vapor transport method \cite{Wen}. The single crystal exhibits a plate-like shape, as shown in the inset of Fig. 1(c). Its large natural surface was determined to be the (001) plane by using an x-ray diffractometer (D8 Advance, Bruker), as illustrated in Fig. 1(c). Magnetization measurements were performed in commercial SQUID and physical property measurement system (PPMS) (Quantum Design). The specific heat was measured in the PPMS by the relaxation method. The $\alpha$-RuCl$_3$ single crystal for the thermal conductivity measurements was cut into a rectangular shape of dimensions 2.01 $\times$ 1.08 mm$^2$ in the $ab$ plane, with a thickness of 0.09 mm along the $c$ axis. Four silver wires were directly attached to the sample with silver paint. The thermal conductivity was measured in a dilution refrigerator, using a standard four-wire steady-state method with two RuO$_2$ chip thermometers, calibrated $in$ $situ$ against a reference RuO$_2$ thermometer. Magnetic fields were applied within the large natural surface (the $ab$ plane) and perpendicular to the heat current.

\begin{figure}
\includegraphics[clip,width=6.6cm]{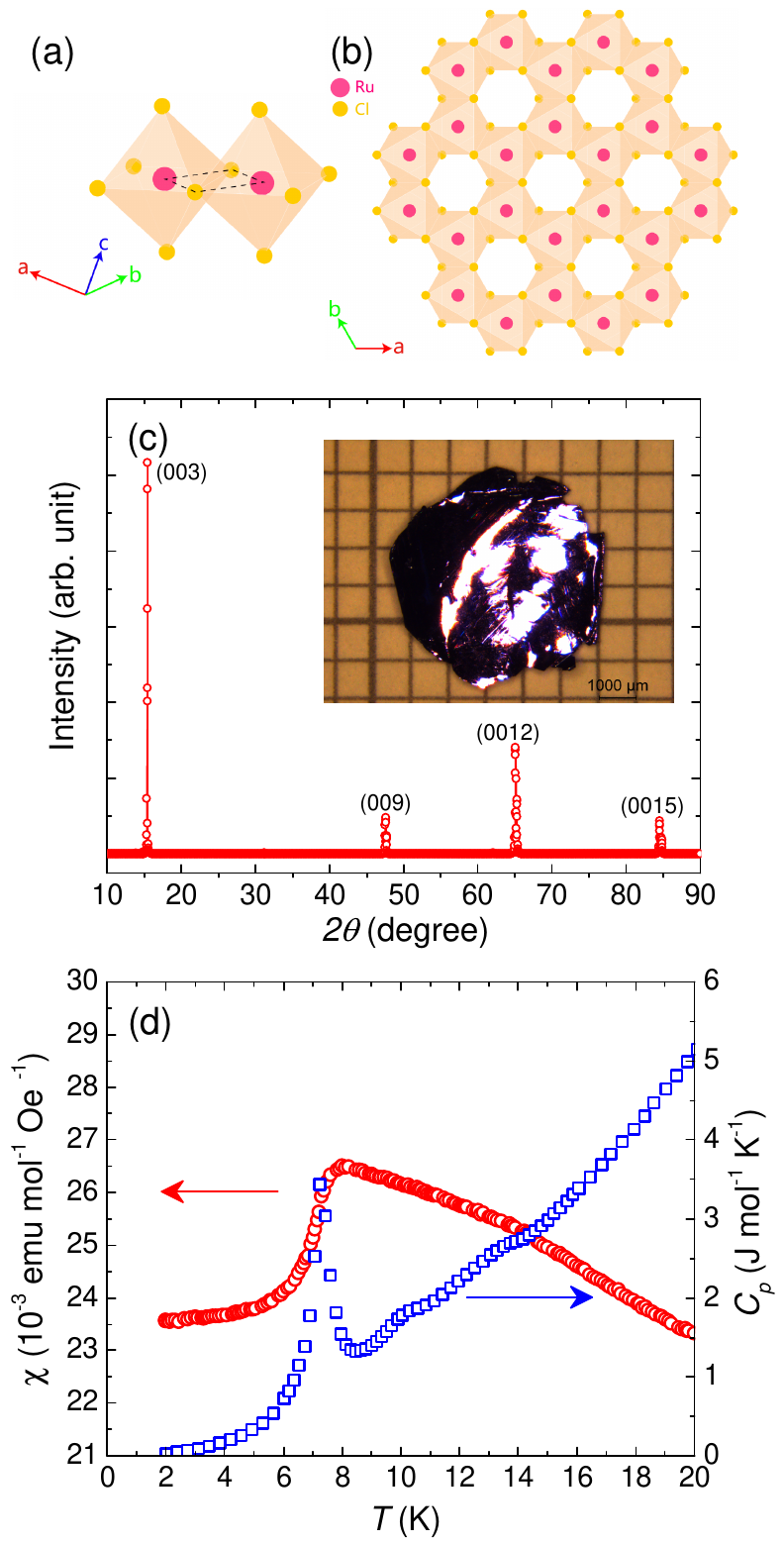}
\caption{(a) Illustration of edge-sharing RuCl$_6$ octahedra that give rise to a dominant Kitaev exchange interactions in $\alpha$-RuCl$_3$. Dashed lines are two Ru-Cl-Ru exchange paths exhibiting a nearly 90$^\circ$ bonding geometry. Ru and Cl atoms are displayed by pink and yellow balls, respectively. (b) The $ab$-plane structure of $\alpha$-RuCl$_3$, where edge-sharing RuCl$_6$ octahedra form a honeycomb network. (c) Room-temperature x-ray diffraction pattern from the large natural surface of the $\alpha$-RuCl$_3$ single crystal. Only (00$l$) Bragg peaks show up, indicating that the large natural surface is $ab$ plane. Inset: the optical image of a typical $\alpha$-RuCl$_3$ single crystal. (d) Temperature dependence of the magnetic susceptibility at $H$ = 0.1 T $\parallel$ $ab$ and the specific heat in zero field for the $\alpha$-RuCl$_3$ single crystal.}
\end{figure}

Figure 1(a) shows the illustration of edge-sharing RuCl$_6$ octahedra in $\alpha$-RuCl$_3$. There exist two Ru-Cl-Ru exchange paths (dashed lines), which exhibit a nearly 90$^\circ$ bonding geometry. Such a superexchange process has proven be crucial for inducing anisotropic Kitaev interactions in strongly spin-orbit coupled compounds \cite{Khaliullin}. The in-plane honeycomb structure of $\alpha$-RuCl$_3$, where edge-sharing RuCl$_6$ octahedra form a honeycomb network, is shown in Fig. 1(b). Temperature dependence of the magnetic susceptibility at $H$ = 0.1 T $\parallel$ $ab$ and the specific heat in zero field for the $\alpha$-RuCl$_3$ single crystal is plotted in Fig. 1(d). It can be clearly seen that there is only one magnetic phase transition at 8 K from $\chi$ and $C_p$. In earlier studies, the single crystals exhibit two magnetic phase transitions at $T_{N1}$ $\sim$ 8 K and $T_{N2}$ $\sim$ 14 K \cite{Sears,Kindo,Baenitz}. The lower (higher) temperature transition is due to the ABC(AB)-type stacking of the honeycomb layers \cite{Nagler}. The single crystals with a single transition at $T_{N1}$ $\sim$ 8 K are ideal for the study of the physics in $\alpha$-RuCl$_3$ \cite{neutron Science,Wen,Buchner,Hess,Buchner1}.

\begin{figure}
\includegraphics[clip,width=6.6cm]{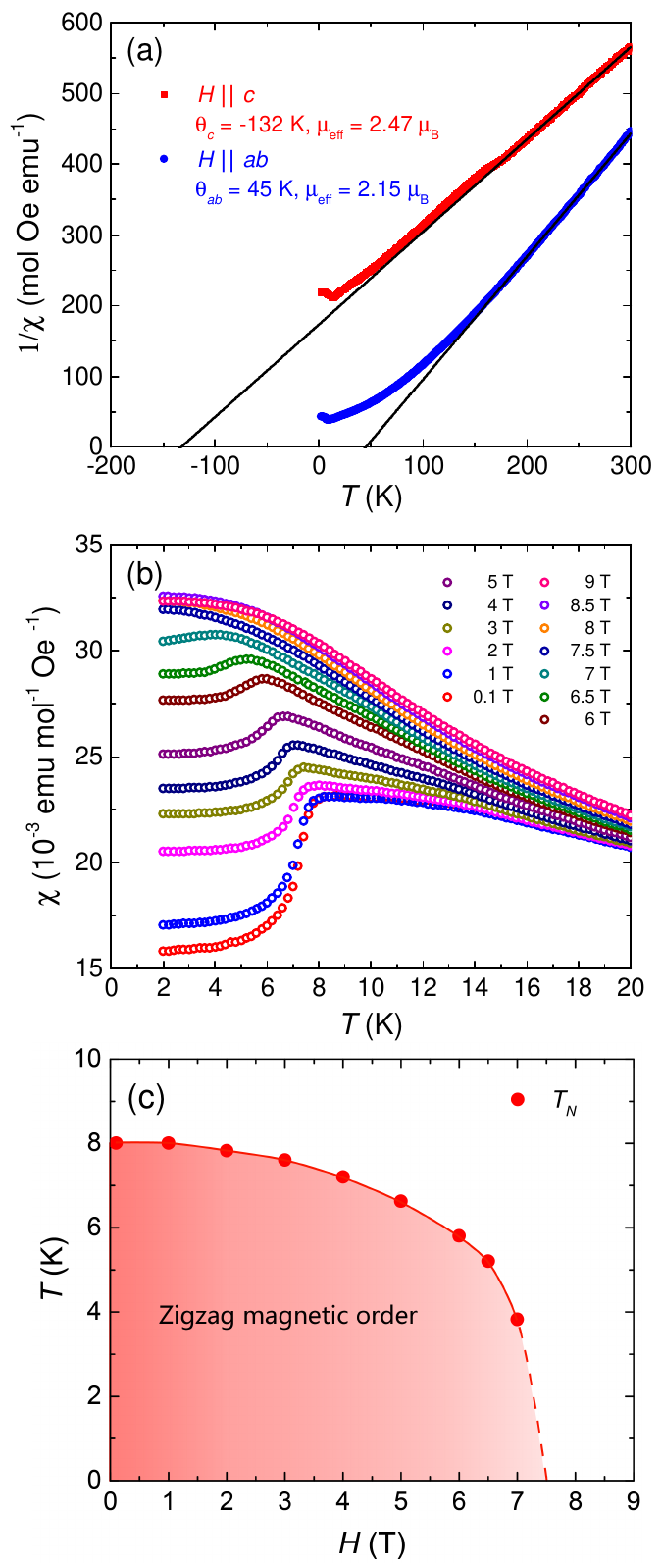}
\caption{(a) Temperature dependence of inverse magnetic susceptibility for the $\alpha$-RuCl$_3$ single crystal in a magnetic field $H$ = 0.1 T applied both parallel and perpendicular to the $ab$ plane. The black lines are fits to the Curie-Weiss law. (b) The magnetic susceptibility of the $\alpha$-RuCl$_3$ single crystal in various magnetic fields up to 9 T ($H$ $\parallel$ $ab$). (c) A schematic $T$-$H$ phase diagram for $\alpha$-RuCl$_3$, where $T_N$ is obtained from the kink in the $\chi$$(T)$ curve. The zigzag magnetic order disappears at a critical field $H_c \sim$ 7.5 T.}
\end{figure}

Figure 2(a) presents the temperature dependence of the inverse susceptibility for the $\alpha$-RuCl$_3$ single crystal in a magnetic field $H$ = 0.1 T applied both parallel and perpendicular to the $ab$ plane. It is highly anisotropic ($\chi_{ab}$/$\chi_{c}$ $\approx$ 5.1 at 20 K), which is attributed to the strongly anisotropic $g$ factor \cite{Kindo}. The slight anomaly around 170 K for $H$ $\parallel$ $c$ results from the structural phase transition \cite{Sears,Buchner}. The data above 180 K for the two field directions can be fitted to the Curie-Weiss law. The obtained Curie-Weiss temperatures $\Theta_c$ $\approx$ $-$132 K and $\Theta_{ab}$ $\approx$ 45 K indicate the effective antiferromagnetic and ferromagnetic exchange interactions, respectively. The effective moments obtained from $\chi_c$ and $\chi_{ab}$ are 2.47 $\mu_B$ and 2.15 $\mu_B$, both larger than the spin-only value of 1.73 $\mu_B$ for the low-spin state of Ru$^{3+}$, suggesting a possibly significant contribution from the orbital moment \cite{Sears}. The magnetic susceptibility of the $\alpha$-RuCl$_3$ single crystal in various magnetic fields up to 9 T ($H$ $\parallel$ $ab$) are plotted in Fig. 2(b). With increasing fields, the phase transition gradually shifts towards lower temperatures and eventually disappears above $\sim$7.5 T. These results are summarized in the $T$-$H$ phase diagram, as shown in Fig. 2(c), where the transition temperature $T_N$ is determined by the kink in the $\chi$$(T)$ curve. The heat transport behavior across the field-induced phase transition is the focus of current study.

\begin{figure}
\includegraphics[clip,width=6.16cm]{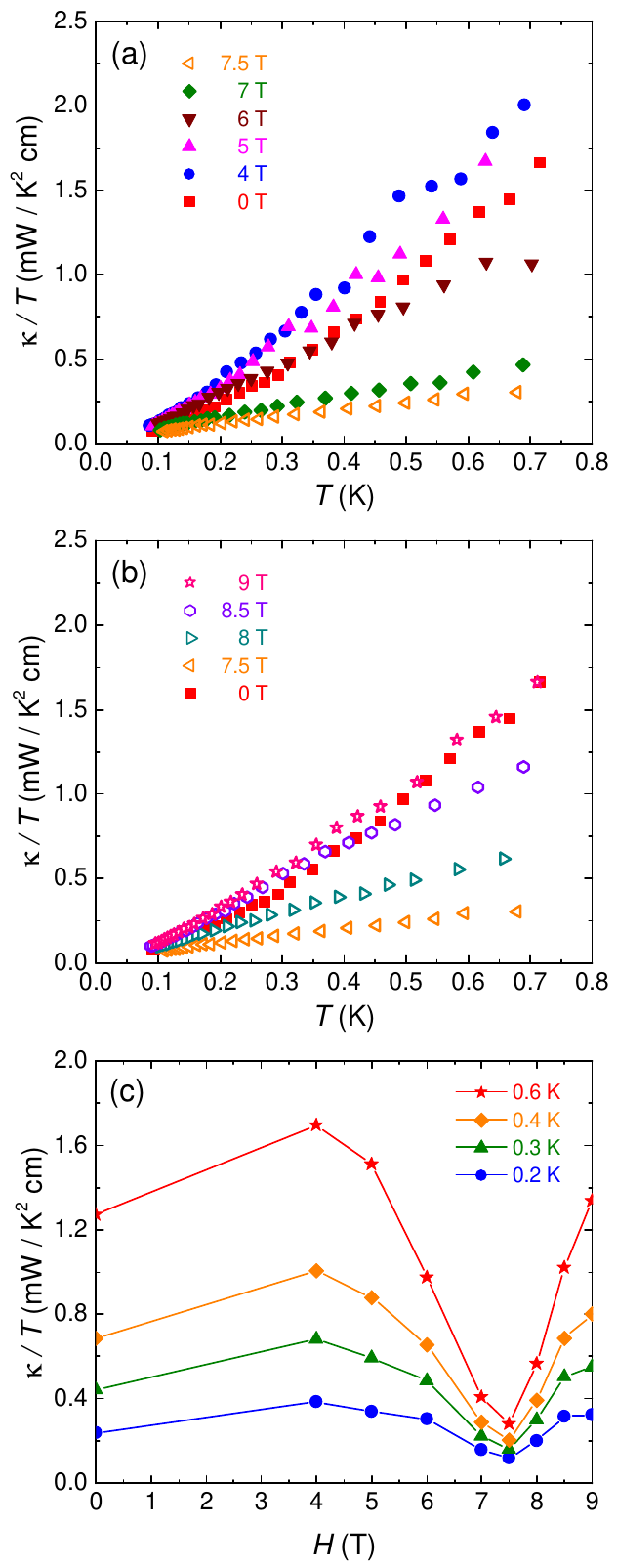}
\caption{(a) and (b) The in-plane thermal conductivity of the $\alpha$-RuCl$_3$ single crystal in zero and finite magnetic fields. (c) Field dependence of the $\kappa/T$ at $T$ = 0.2, 0.3, 0.4, and 0.6 K, respectively. For all curves, the $\kappa/T$ first increases slightly for $H$ $<$ 4 T, then drops rapidly until 7.5 T, followed by a sharp increase. The minimum $\kappa/T$ at $\sim$7.5 T corresponds to $H_c$, where the zigzag magnetic order disappears.}
\end{figure}

Figure 3(a) and 3(b) show the in-plane thermal conductivity of the $\alpha$-RuCl$_3$ single crystal in zero and finite magnetic fields. As the magnetic field is increased, the $\kappa/T$ first increases slightly for $H$ $<$ 4 T, then drops rapidly until 7.5 T, followed by a sharp increase, as illustrated in Fig. 3(c). The minimum of $\kappa/T$ at $\sim$7.5 T corresponds to $H_c$, where the zigzag magnetic order disappears. Such a minimum of $\kappa(H)$ was also observed in previous thermal conductivity experiments \cite{Hess,Lee}, which likely results from the strong scattering of phonons by magnetic fluctuations at the critical point. The sharp increase of $\kappa(H)$ above $H_c$ was explained in terms of two different scenarios \cite{Hess,Lee}. In Ref. \cite{Lee}, the linear rise of the $\kappa$ is interpreted as the contribution from the gapless PKE. However, based on the similar behavior of $\kappa_{ab}(H)$ and $\kappa_c(H)$, Hentrich $et$ $al.$ argued that the $\kappa$ of $\alpha$-RuCl$_3$ is purely contributed by phonons, and the magnetic excitations do not contribute to $\kappa$ but can scatter phonons strongly \cite{Hess}. With increasing fields above $H_c$, the magnetic excitations are increasingly gapped out, thus reducing the scattering, leading to the enhancement of phonon thermal conductivity $\kappa$ \cite{Hess}. To clearly examine whether there is contribution to $\kappa$ from the gapless magnetic excitations, it is essential to know the asymptotic behaviors of $\kappa/T$ at $T$ $\rightarrow$ 0, reflecting the nature of the ground state.

\begin{figure}
\includegraphics[clip,width=6.6cm]{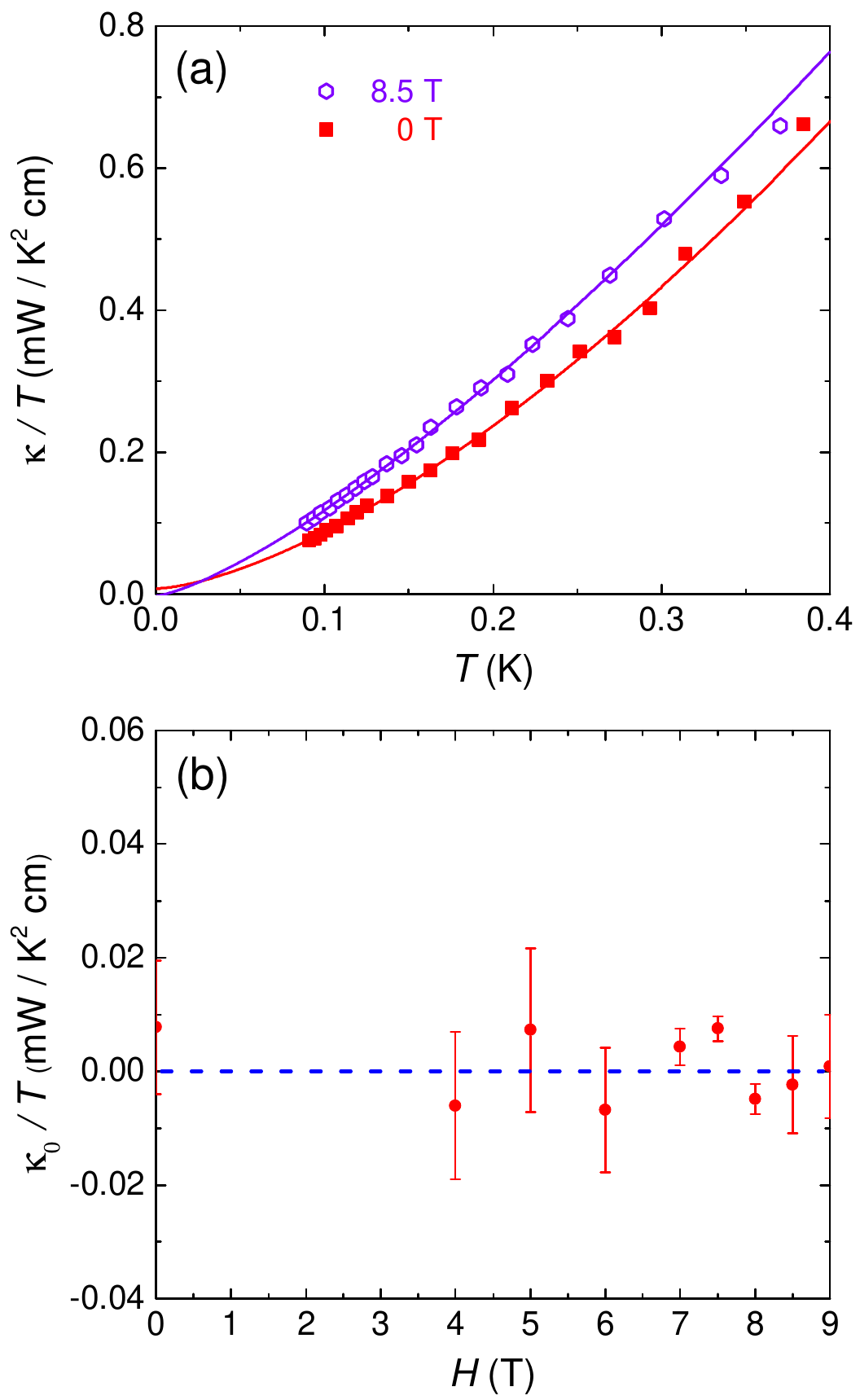}
\caption{(a) The fits of the thermal conductivity data of the $\alpha$-RuCl$_3$ single crystal in the zigzag magnetic order ($H$ = 0 T) and the high-field phase ($H$ = 8.5 T) to $\kappa/T$ = $a$ + $bT^{\alpha-1}$ below 0.4 K, represented by the solid lines. (b) Field dependence of the residual linear term $\kappa_0/T$. These values of $\kappa_0/T$ are virtually zero, indicating the absence of gapless fermionic magnetic excitations.}
\end{figure}

In Fig. 4(a), we first fit the zero-field data below 0.4 K to $\kappa/T$ = $a$ + $bT^{\alpha-1}$, in which the two terms $aT$ and $bT^{\alpha}$ represent contributions from itinerant fermionic excitations (if they exist) and phonons, respectively \cite{Takagi,Taillefer}. Because of the specular reflections of phonons at the sample surfaces, the power $\alpha$ in the second term is typically between 2 and 3 \cite{Takagi,Taillefer}. The fitting gives $\kappa_0/T$ $\equiv$ $a$ = 0.007 $\pm$ 0.011 mW K$^{-2}$ cm$^{-1}$ and $\alpha$ = 2.52 $\pm$ 0.09. Considering our experimental error bar $\pm$ 0.005 mW K$^{-2}$ cm$^{-1}$, the $\kappa_0/T$ at zero field is virtually zero. Such a pure phonon thermal conductivity at zero field is reasonable, since $\alpha$-RuCl$_3$ is an insulator with a magnon gap in the zigzag magnetic state \cite{Nagler,Wen}. Recently, a theoretical calculation relevant to $\alpha$-RuCl$_3$ based on the quantum Monte Carlo simulation found that itinerant Majorana fermions from the fractionalization of quantum spins can carry heat and contribute a finite residual linear term $\kappa_0/T$ \cite{Motome}. Therefore, if the magnetic excitations are gapless in the high-field phase of $\alpha$-RuCl$_3$, e.g., the PKE suggested in Ref. \cite{Lee}, one may expect a finite residual linear term $\kappa_0/T$. However, by fitting the 8.5 T data to $\kappa/T$ = $a$ + $bT^{\alpha-1}$, we obtain $\kappa_0/T$ = -0.002 $\pm$ 0.009 mW K$^{-2}$ cm$^{-1}$ and $\alpha$ = 2.33 $\pm$ 0.05, as in Fig. 4(a). Furthermore, we plot the field dependence of $\kappa_0/T$ across the field-induced phase transition in Fig. 4(b). All of these values are found to be negligible in our field range. Therefore, the absence of $\kappa_0/T$ above the critical field $H_c$ demonstrates the lack of gapless magnetic excitations in the high-field phase of $\alpha$-RuCl$_3$, such as massless Majorana fermions \cite{Lee}. In other words, our ultralow-temperature thermal conductivity measurements do not support the gapless scenarios of the high-field phase in $\alpha$-RuCl$_3$.

Indeed, a field-induced gapped QSL state has been predicted in $\alpha$-RuCl$_3$ in the field range of 8 $\sim$ 15 T by exact diagonalization and density-matrix renormalization group calculations for extended Kitaev-Heisenberg spin Hamiltonians \cite{Hozoi}. Experimentally, a field-induced gap opening has also been reported in several experiments including inelastic neutron scattering \cite{Nagler1}, nuclear magnetic resonance \cite{Buchner,Klanjsek}, specific heat \cite{Kim,Buchner1}, and thermal conductivity measurements \cite{Hess}. The enhancement of thermal conductivity in the high-field phase may be explained by the increasing gap scenario \cite{Hess}.

We would like to discuss the gap amplitude near the critical field $H_c$. The results of a previous thermal conductivity study and a specific heat study suggest that the loss of magnetic order at $H_c$ is accompanied by the closing of the magnetic excitation gap \cite{Hess,Buchner1}. In that case, there should be direct contribution to $\kappa_0/T$ from the gapless magnetic excitations at $H_c$, which is not observed in our data. This indicates that either (a) the gap is still finite at $H_c$, as suggested by another specific heat study with the extrapolation of the gap from high fields \cite{Kim}; or (b) the gap vanishes around $H_c$, i.e., the gapless magnetic excitations do exist, but are localized so that they cannot conduct heat, possibly owing to magnetic defects \cite{Buchner} or crystallographic domains \cite{Buchner1}.

Note that there is also a more general theoretical scenario, ascribing the continuum observed by inelastic neutron scattering in $\alpha$-RuCl$_3$ to incoherent excitations originating from strong magnetic anharmoniticity \cite{Valenti}, instead of the most discussed explanation referring to a coherent continuum of fractional excitations analogous to the celebrated Kitaev spin-liquid. We are not clear what the thermal conductivity behavior is if this more general scenario is the case in $\alpha$-RuCl$_3$.

In summary, we have measured the thermal conductivity of a $\alpha$-RuCl$_3$ single crystal down to 80 mK and up to 9 T. The field dependence of the thermal conductivity exhibits a minimum around 7.5 T, where the field-induced phase transition takes place. The extrapolation of the thermal conductivity down to zero temperature reveals no significant contribution from itinerant gapless fermionic excitations in the high-field phase. These results impose clear constraints on the nature of the high-field phase and are also expected to help distinguish between theoretical scenarios proposed recently.

This work is supported by the Ministry of Science and Technology of China (Grant Nos. 2015CB921401 and 2016YFA0300503), the Natural Science Foundation of China (Grant Nos. 11422429, 11374143, and 11674157), the NSAF (Grant No. U1630248), the Program for Professor of Special Appointment (Eastern Scholar) at Shanghai Institutions of Higher Learning, and STCSM of China (No. 15XD1500200).\\

\noindent $^\dag$ Y. J. Yu, Y. Xu, and K. J. Ran contributed equally to this work.\\
\noindent $^\ddag$ E-mail: jwen$@$nju.edu.cn\\
\noindent $^*$ E-mail: shiyan$\_$li$@$fudan.edu.cn

\end{document}